# Superconducting and normal-state anisotropy of the doped topological insulator $Sr_{0.1}Bi_2Se_3$

M. P. Smylie[1,2,*], K. Willa[1], H. Claus[1], A. E. Koshelev[1], K. W. Song[1], W.-K. Kwok[1], Z. Islam[3], G. D. Gu[4], J. A. Schneeloch[4,5], R. D. Zhong[4,6], and U. Welp[1]

[1]Materials Science Division, Argonne National Laboratory, 9700 S. Cass Ave, Lemont, Illinois, 60439, USA

[2]Department of Physics, University of Notre Dame, Notre Dame, Indiana, 46556, USA

[3]Advanced Photon Source, Argonne National Laboratory, 9700 S. Cass Ave, Lemont, Illinois, 60439, USA

[4]Condensed Matter Physics and Materials Science Department, Brookhaven National Laboratory, Upton, New York, 11793, USA

[5]Department of Physics and Astronomy, Stony Brook University, Stony Brook, New York, 11794, USA

[6]Department of Materials Science and Engineering, Stony Brook University, Stony Brook, New York, 11794, USA

[*]msmylie@anl.gov

**ABSTRACT**

$Sr_xBi_2Se_3$ and the related compounds $Cu_xBi_2Se_3$ and $Nb_xBi_2Se_3$ have attracted considerable interest, as these materials may be realizations of unconventional topological superconductors. Superconductivity with $T_c \sim 3$ K in $Sr_xBi_2Se_3$ arises upon intercalation of Sr into the layered topological insulator $Bi_2Se_3$. Here we elucidate the anisotropy of the normal and superconducting state of $Sr_{0.1}Bi_2Se_3$ with angular dependent magnetotransport and thermodynamic measurements. High resolution x-ray diffraction studies underline the high crystalline quality of the samples. We demonstrate that the normal state electronic and magnetic properties of $Sr_{0.1}Bi_2Se_3$ are isotropic in the basal plane while we observe a large two-fold in-plane anisotropy of the upper critical field in the superconducting state. Our results support the recently proposed odd-parity nematic state characterized by a nodal gap of $E_u$ symmetry in $Sr_xBi_2Se_3$.

## Introduction

Following the discovery of topological insulators[1,2], the search for a superconducting analogue of a topological insulator has gained considerable interest in the condensed matter physics community. A topological superconductor (TSC)[3–5] has a bulk superconducting energy gap (nodal or nodeless)[6] but has gapless surface states which are of great interest both for fundamental physics, as they can host Majorana quasiparticles[7], and also for applied physics, as the non-Abelian statistics of surface-state excitations have important implications for robust quantum computing[8–10].

The topological nature of the superconducting state is determined by the symmetry of the superconducting order parameter and the shape of the Fermi surface. In a time-reversal and inversion symmetric system, odd-parity pairing, where $\Delta(-\mathbf{k}) = -\Delta(\mathbf{k})$, and a Fermi surface that contains an odd number of time-reversal invariant momenta are necessary requirements[5]. In materials with weak spin-orbit coupling, odd-parity pairing corresponds to spin-triplet pairing; for certain strong spin-orbit coupling systems, unique unconventional superconducting states are possible[11] that may qualify as topological superconductivity. Currently, two paths towards topological superconductivity are being investigated: proximity-induced TSC[7,12,13] at the interface between a conventional superconductor and a topological insulator or a strong spin-orbit coupled semiconductor, respectively, and via chemical doping of bulk topological insulators. Of the superconducting doped topological insulators, the $M_xBi_2Se_3$ family of materials (M = Cu, Nb, Sr)[14–16] has generated the most interest as high quality, mm-scale single crystals are available. Topological order observed via ARPES measurements[17], and magnetization measurements[18,19] is consistent with a spin-triplet pairing state. Calorimetry measurements[20] are not in full agreement with conventional BCS theory, and low-temperature penetration depth measurements[21] indicate nodes in the superconducting energy gap. The observation of zero-bias conductivity peaks in point-contact spectroscopy measurements[22–24] has been interpreted as evidence for Majorana surface states.

The $M_x$Bi$_2$Se$_3$ family maintains the trigonal $R\bar{3}m$ structure of the parent compound, which makes recent observations[25–29] of twofold symmetry in several quantities below $T_c$ in the $M_x$Bi$_2$Se$_3$ family of materials all the more surprising. A nematic superconducting state with a two-component order parameter has been proposed[30–32] to explain these results. This state has $E_u$ symmetry and odd-parity pairing, and allows for states with complete, albeit anisotropic, superconducting gap as well as for a gap with point nodes. Despite the unconventional nature, the superconducting state has been shown to be robust against disorder scattering[33–35]. Fig. 1(a) shows the $R\bar{3}m$ crystal structure of Sr$_{0.1}$Bi$_2$Se$_3$, the same as that of the parent compound Bi$_2$Se$_3$ with a slightly extended $c$ axis due to intercalation of the Sr atom in the gap between adjacent quintuple layers of Bi$_2$Se$_3$[36] while Fig. 1(b) shows the threefold symmetric basal plane, with the $a$ (blue) and $a^*$ (pink) directions marked by arrows. Fig. 1(c) shows the proposed twofold symmetric $\Delta_4$ superconducting gap structure, which breaks crystallographic rotational symmetry in the basal plane[30].

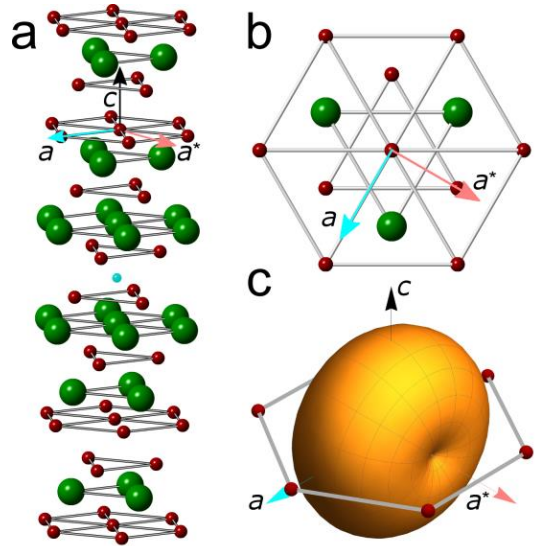

**Figure 1.** Fig. 1: (a) Crystal structure of Sr$_x$Bi$_2$Se$_3$, with directions $a$ (blue), $a^*$ (pink) and $c$ (black) marked. The dopant Sr ion (blue) sits in the van der Waals gap between quintuple layers of Bi (green) and Se (red) ions. (b) The threefold symmetric basal plane. (c) The two-dimensional $\Delta_4$ superconducting gap structure has basis functions ($\Delta_{4x}$) with nodes on the mirror plane and ($\Delta_{4y}$) with deep minima in the perpendicular direction. This gap breaks the threefold crystal symmetry and gives rise to the nematic state with twofold symmetry.

Here, we present the first thermodynamic determination of the anisotropy of the upper critical field of Sr$_{0.1}$Bi$_2$Se$_3$ single crystals through measurements of the temperature dependence of the reversible magnetization, in addition to angular-dependent magnetotransport measurements. Both quantities yield a large twofold in-plane anisotropy of $H_{c2}$ in which the high-$H_{c2}$ direction is aligned with the $a$-axis of the crystal structure. We find that the normal state resistivity of Sr$_{0.1}$Bi$_2$Se$_3$ is isotropic in pairs of samples cut at 90° from the same starting crystal, which excludes conventional mass anisotropy from being the cause of the anisotropy in $H_{c2}$. Furthermore, temperature dependent measurements of the normal-state magnetization show that Sr$_{0.1}$Bi$_2$Se$_3$ is diamagnetic with an isotropic susceptibility of $\sim -2\cdot 10^{-6}$ (CGS) which largely originates from the core diamagnetism. X-ray diffraction studies indicate that the extinction rule for the $R\bar{3}m$ crystal structure is fulfilled to a level of $10^{-6}$ implying that deviations from the ideal $R\bar{3}m$ structure are exceedingly small. We thus conclude that the origin of the twofold anisotropy of the superconducting properties is likely caused by an anisotropic gap structure consistent with the nematic $E_u$ state.

## Results

We present results on a series of Sr$_x$Bi$_2$Se$_3$ crystals. Bar-shaped crystals #1a and 1b were cut from the same starting piece oriented at 90° with respect to each other. Field-angle dependent resistivity measurements (Fig. 2) reveal that the nematic state is not tied to the current flow direction. Detailed resistivity measurements as function of applied magnetic field, field orientation and temperature (Figs. 3, 4) on crystal #2 yield the anisotropic phase diagram while high-resolution x-ray diffraction on the same sample (Fig. 7) reveals high crystalline quality. On crystal #3 we performed magnetic measurements of the nematic state (Fig. 5) and of the normal state susceptibility (6). Bar-shaped crystals #4a and 4b were cut from the same starting piece such that they are oriented parallel and perpendicular to the nematic axis, respectively. Finally, we determined the in-plane superconducting anisotropy using magnetization and resistivity measurements on crystal #5, shown in Supplemental Materials.

The resistivity as a function of in-plane angle in an applied magnetic field of 1 T is shown in Fig. 2 for $Sr_{0.1}Bi_2Se_3$ crystals #1a and #1b at temperatures ranging from 1.7 K (black) to 2.9 K (purple) in increments of 0.1 K. The two crystals were cut from the same starting material at 90° with respect to each other (see inset of Fig. 6). Here, the crystals were intentionally cut such that the long axes did not lie along or perpendicular to the nematic axis. 0° (red arrow) marks the direction of the current ($I$ = 0.1 mA). The angular dependence of the resistivity thus reflects the angular dependence of the upper critical field $H_{c2}$, as directions with higher $H_{c2}$ will remain superconducting whereas directions with lower $H_{c2}$ will be resistive at a fixed temperature. Twofold anisotropy is evident in both Fig. 2(a) and Fig. 2(b). As temperature is increased from base temperature through the superconducting transition, the twofold anisotropy is eventually lifted, reaching an angle independent normal state. There is an obvious 90° rotation in the axes of high and low $H_{c2}$ between Fig. 2(a) and Fig. 2(b), behavior that was seen in all measured sets of crosscut crystals, coming from the 90° rotation of the crystalline axes in cut crystal pairs. These data demonstrate that the observed twofold anisotropy is tied to the crystal structure, and is not an effect due to current flow, such as Lorentz force driven vortex motion.

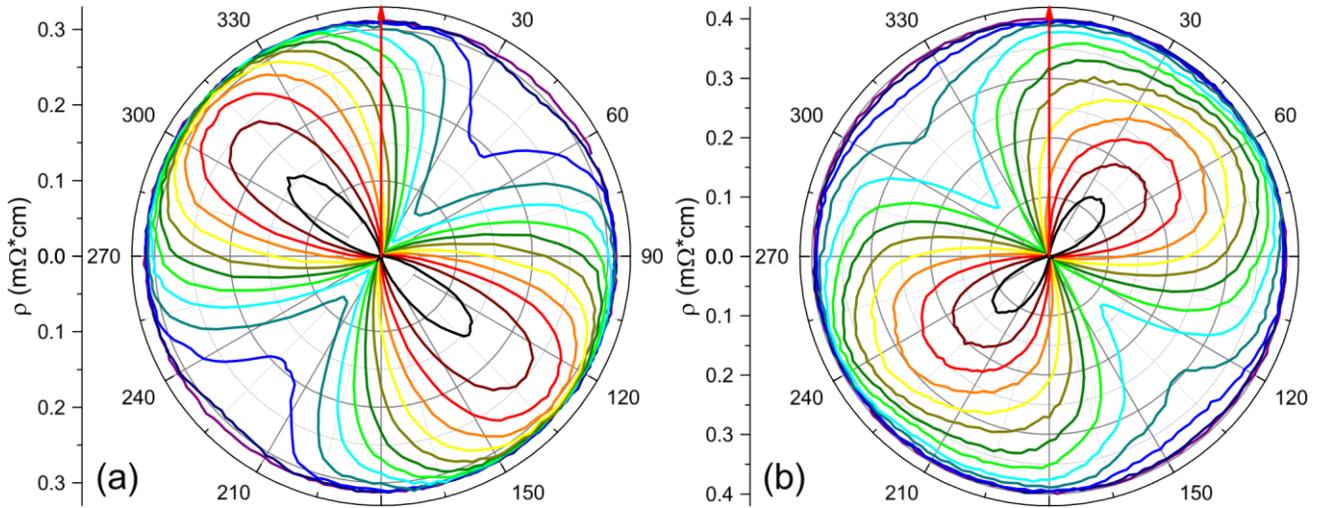

**Figure 2.** R(θ) in an applied magnetic field of $\mu_0|H_{ab}|$ = 1 T in two crystals of $Sr_{0.1}Bi_2Se_3$ cut at 90° relative to each other from a single larger crystal in the temperature range 1.7 K (black) to 2.9 K (purple) in 0.1 K steps; $I$ = 0.1 mA. 0° marks the direction of current (red arrow). As temperature increases, the twofold nematic symmetry becomes rotationally isotropic. A 90° rotation of the crystalline axis with identical directions of current (red arrows) results in the nematic axis rotating by 90°; vortex motion depends on the relative directions of $I$ and $B$ and would be unchanged under only crystalline rotation.

To further investigate the angular anisotropy of $H_{c2}$, a series of ρ(T) curves were measured on crystal #2 with $T_c \approx$ 2.9 K in different applied magnetic fields with the field vector along the directions of maximum and minimum in-plane $H_{c2}$ as well as along the c axis of the crystal [Fig. 3(a), (b), (c)]. X-ray diffraction on this sample (see Fig. 7) reveals that the directions of high (low) in-plane $H_{c2}$ correspond to the crystallographic a and a* directions, respectively (see Fig. 1), consistent with previous reports[26,37]. Figures 3(a), (b), (c) show that on increasing field the transitions stay sharp and shift uniformly to lower temperatures. A weak normal-state magnetoresistance is observed only for H//c. Figure 3(d) shows the magnetic phase diagram along the principal axes with $T_c$ taken as the midpoint of the resistive transitions. The in-plane anisotropy Γ is ~ 4.5. Reported values for the anisotropy (also determined from the resistive midpoints) range from 6.8 for nominal 10% doping to 2.7 for nominal 15% doping, both at 1.9 K[26], whereas on samples with unspecified doping levels an in-plane anisotropy value of $\sim 2.8$ was obtained[37]. In the temperature range covered here, $\Gamma$ is approximately temperature-independent.

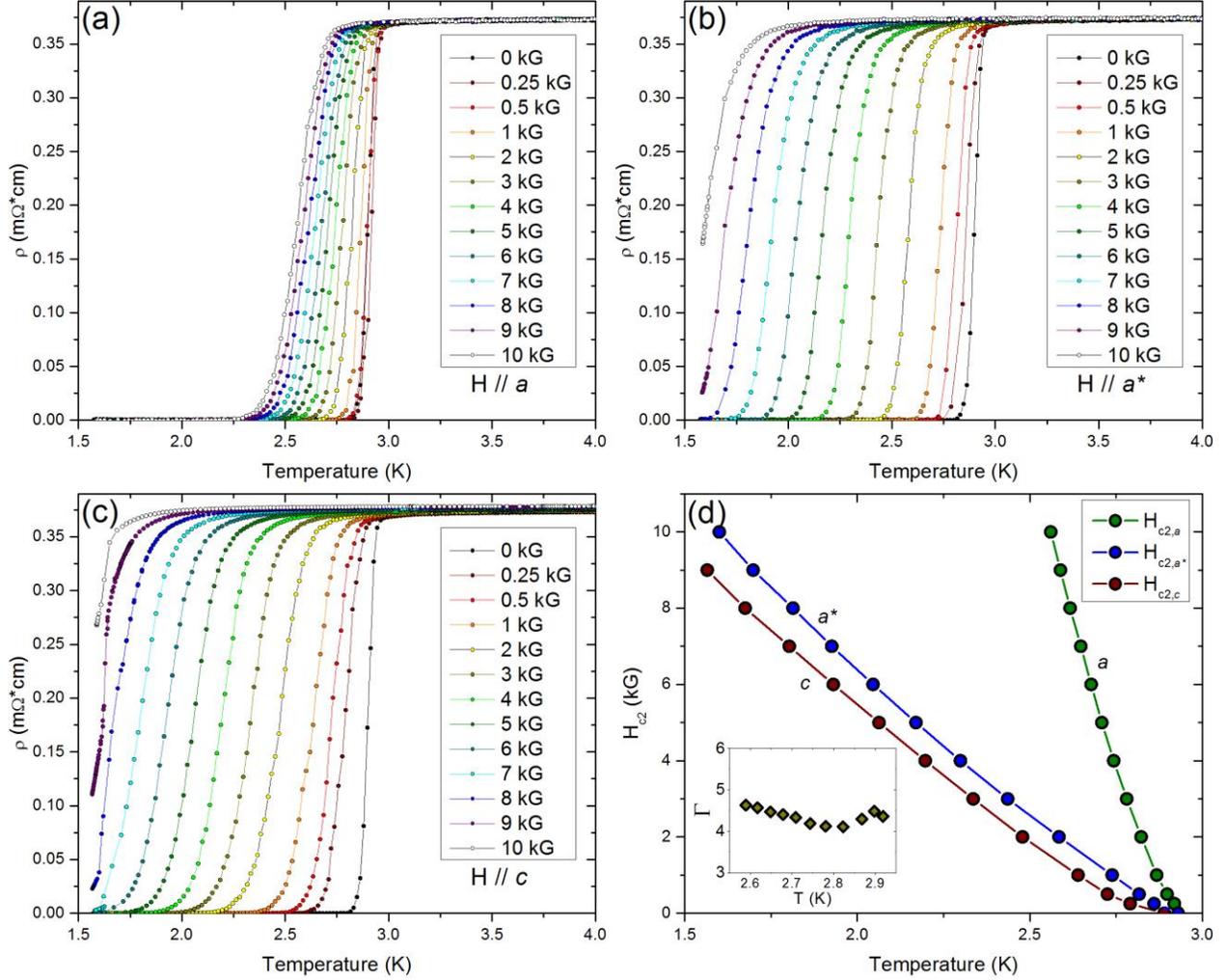

**Figure 3.** $\rho(T)$ of $Sr_{0.1}Bi_2Se_3$ crystal #2 in increasing magnetic field with the field vector in different orientations. (a) Field vector $H//a$. (b) Field vector $H//a*$. (c) Field vector $H//c$. (d) Magnetic phase diagram. There is a large anisotropy of $\sim 4$ between the two in-plane orientations which are 90° apart. The inset shows the in-plane $\Gamma = H_{c2}^a / H_{c2}^{a*}$.

In standard single-band Ginzburg-Landau (GL) theory the anisotropy of $H_{c2}$ is given by the anisotropy of the effective masses[38]

$$\frac{H_{c2}^{(i)}}{H_{c2}^{(j)}} = \sqrt{\frac{m_j}{m_i}} = \Gamma > 1 \tag{1}$$

where $m_i$ and $m_j$ are the effective masses along the principal crystal directions $i$ and $j$. The unit vectors $i, j, k$ define a Cartesian coordinate system; here, $i = a$, $j = a*$, $k = c$. The angular variation of $H_{c2}$ in the $ij$-plane is then given by[38]

$$H_{c2}(T,\theta) = \frac{H_{c2}^{(j)}}{\sqrt{cos^2(\theta)+\Gamma^{-2}sin^2(\theta)}} \qquad (2)$$

where $\theta$ is measured from the $j$-direction (low $H_{c2}$). Approximating the $H_{c2}$-line as linear, the angular dependence of $T_c(H,\theta)$ in a given field $H$ can be obtained from Eq. 2 as

$$T_c(H,\theta) = T_{c0} + \frac{H}{\partial H_{c2}^{(j)}(T)/\partial T}\sqrt{cos^2(\theta)+\Gamma^{-2}sin^2(\theta)} \qquad (3)$$

Data in a field of $\mu_0 H$ = 1 T are obtained from the polar diagram of $\rho(T,\theta)$ [(Fig. 4(a)] by tracing for which values of $T$ and $\theta$ the resistivity crosses the 50% value. The results for $T_c(\theta)$ are shown in Fig. 4(b) together with a fit to Eq. 3. The fit yields an in-plane anisotropy of $\Gamma \sim 3.8$, in reasonable agreement with the data in Fig. 3(d). The small difference in anisotropy may arise from deviations from linearity of the phase boundaries.

We obtain the first thermodynamic measurement of the in-plane anisotropy of the upper critical field of $Sr_{0.1}Bi_2Se_3$ from the temperature dependence of the magnetization of crystal #3 with $T_c \approx$ 3 K. Figure 5 shows data taken in a field of 0.4 T applied along the high and low $H_{c2}$-directions, respectively. A shift in $T_c$, defined as the intersection of a linear fit to the $M(T)$-data with the $M$ = 0 line, and a change in the slope $dM/dT$ with field angle are clearly seen. The inset of Fig. 5 displays the twofold symmetric angular variation of the slope $dM/dT$ in which a low value of the slope corresponds to a high value of $T_c$. Such behavior is expected in conventional single-band GL theory of anisotropic superconductors, for which the slope in field direction $i$ is given as

$$\frac{\partial M^{(i)}}{\partial T} = -\frac{1}{8\pi\beta_A(\kappa^{(i)})^2}\frac{\partial H_{c2}^{(i)}}{\partial T} \qquad (4)$$

where $H_{c2}^{(i)} = \phi_0/(2\pi\xi_j\xi_k)$ and $\kappa^{(i)} = \sqrt{(\lambda_j\lambda_k)/(\xi_j\xi_k)} \gg 1$ are the upper critical field and Ginzburg-Landau parameter in direction $i$, respectively, and $\beta_A$ = 1.16 is the Abrikosov number. With the $T$-linear variation of $H_{c2}$ near $T_c$ one finds that the ratio of the slopes for the high-$H_{c2}$ and low-$H_{c2}$ directions is given by the inverse anisotropy

$$\frac{\partial M^{(i)}}{\partial T}\bigg/\frac{\partial M^{(j)}}{\partial T} = \frac{1}{\Gamma} < 1 \qquad (5)$$

Thus, the data shown in the inset of Fig. 5 indicate an anisotropy of $\Gamma \sim 2$ which is smaller than the value deduced from the resistivity measurements (Fig. 3). Data such as shown in Fig. 5(a) taken over the entire angular range in fields of 0.4 T and 0.6 T yield the angular dependence of $T_c$ as shown in Fig. 5(b) for 0.4 T (blue) and 0.6 T (red). Although there is sizable scatter in the data [the error bars in Fig. 5(b) reflect the scatter in $T_c$ obtained on repeated runs], a twofold angular symmetry in this thermodynamic determination of $T_c$ is clearly seen consistent with the twofold symmetry observed in magnetotransport measurements. The data shown in Fig. 5(a) also demonstrate that the superconductivity observed in our $Sr_{0.1}Bi_2Se_3$ crystals is a bulk phenomenon and not filamentary.

Although transport and magnetization measurements yield similar qualitative features of the superconducting phase diagram of $Sr_xBi_2Se_3$, i.e. a sizable in-plane anisotropy, there are clear quantitative differences in the value of the anisotropy deduced from both techniques. Generally, such differences may arise since magnetization and resistivity represent different quantities, the expectation value of the magnitude-squared of the superconducting order parameter and the onset of phase coherence across the sample, respectively. Furthermore, the resistively determined phase boundaries depend on the resistivity criterion used; here we use the 50% criterion. Nevertheless, considering that the resistive transitions shown in Fig. 3 appear 'well-behaved', a difference in anisotropy by a factor of $\sim$ 2 is surprising. In order to rule out the doping-dependence as a cause of the difference in anisotropy seen in magnetization and magnetoresistance measurements, we performed detailed magnetization and resistivity measurements on a large single crystal, sample #5, shown in Supplemental Materials, and reproduce the result that the magnetically determined in-plane superconducting anisotropy is smaller than the resistive result: $\Gamma \sim$ 2.6 versus 5. The reasons for this unexpected behavior are not understood at present, and may be related to the unusual positive curvature observed in $H_{c2}$ in all samples as determined by magnetotransport, or to the existence of surface states which may have different superconducting properties[27] than the bulk.

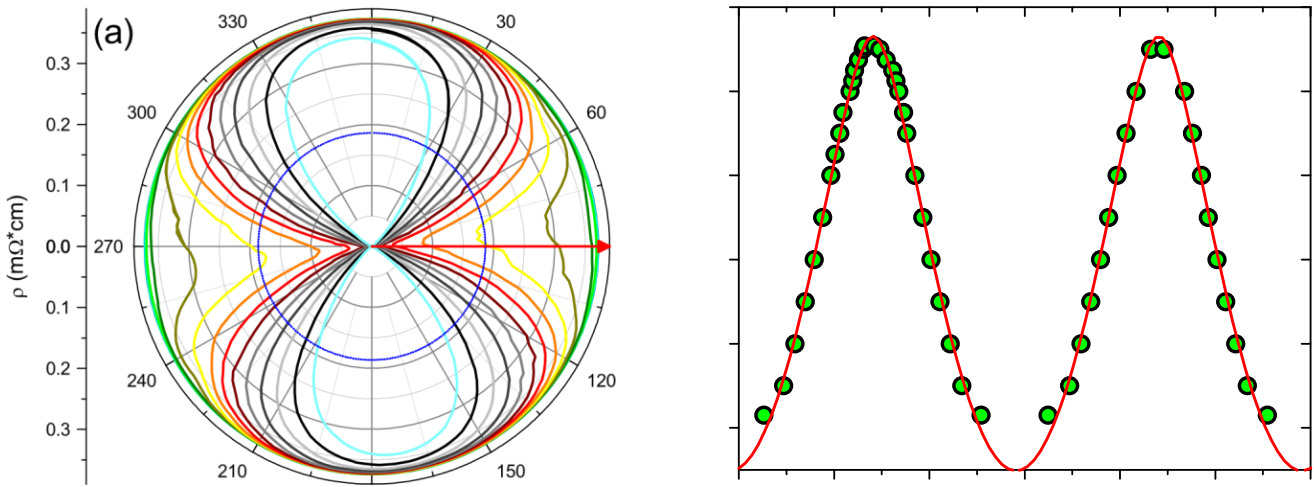

**Figure 4.** (a) $R(\theta)$ for $Sr_{0.1}Bi_2Se_3$ crystal #2 with an in-plane field of 1 T in 0.1 K increments, from 1.7 K (cyan) to 3 K (green). The red arrow marks the direction of current. (b) $T_c(\theta)$ extrapolated from $R(\theta)$ data for the same crystal, taken as where $R(\theta)$ is half the normal-state value, represented by the blue circle in (a). Several additional small $\theta$ windows were measured at multiple temperatures to increase data density. The red line is a fit to the data following the Ginzburg-Landau effective mass formula (see text) yielding $\Gamma \approx 3.8$.

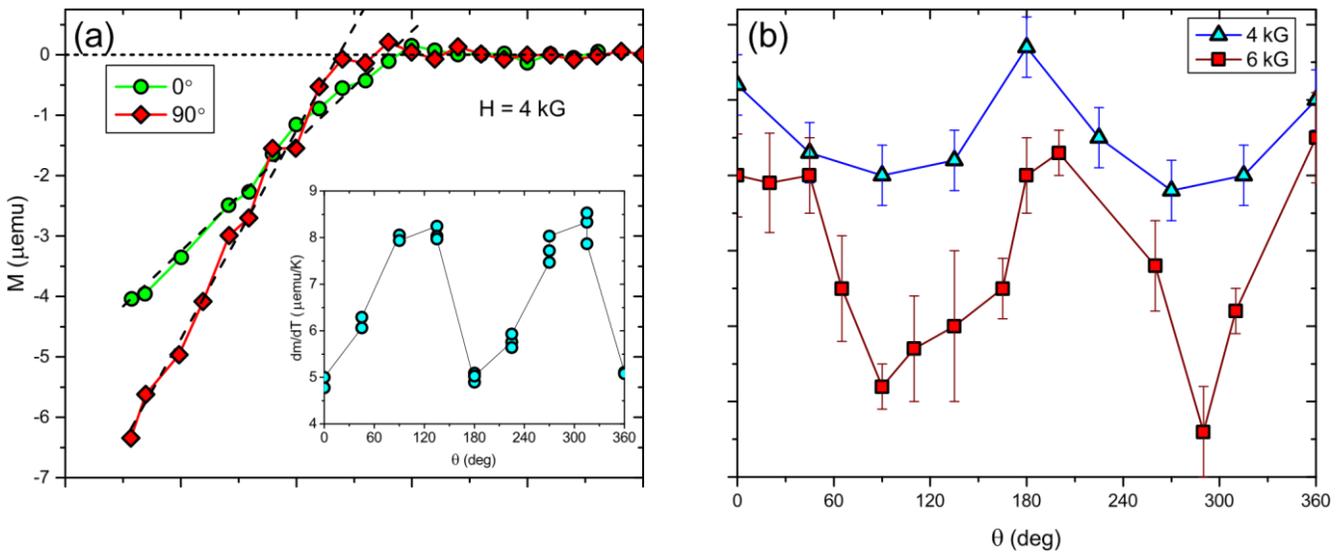

**Figure 5.** (a) $M(T)$ curves as measured by dc SQUID magnetometry on $Sr_{0.1}Bi_2Se_3$ crystal #3 in 4 kG for two different orientations of field 90° apart in the basal plane, with linear fits (dashed lines) below $T_c$. There is a clear difference in $T_c$ taken as where a linear fit of the response (dashed lines) crosses zero. The inset shows the slope of the linear fit vs angle for multiple measurements. A twofold symmetry in $T_c$ is evident. (b) $T_c(\theta)$ with $\theta$ the orientation of magnetic field in the basal plane as measured by dc SQUID magnetometry on a single crystal of $Sr_{0.1}Bi_2Se_3$ in fields of 4 kG (blue) and 6 kG (red). In both fields, $T_c$ is again twofold symmetric.

Figure 6(a) shows the temperature dependence of the resistivity of cross-cut crystals with very sharp superconducting transitions at an onset temperature of ∼ 3.0 K oriented such that the current in crystal #4a flows along the

*a*-direction and in crystal #4b along *a\**, respectively. The anisotropy in the normal-state resistivity is small, < 10%. We note that absolute values of the resistivity have an uncertainty of ~ ±15% due to uncertainties in the dimensions of the samples and contact geometry. At the same time, the upper critical field displays a sizable in-plane anisotropy as expressed by the ratio of the effective masses (Eq. 1). For superconductors with essentially isotropic gaps, these effective masses are the same as those entering the normal state conductivity. Our observed sizable in-plane $H_{c2}$-anisotropy would imply an in-plane resistivity anisotropy of more than 4, which is clearly not consistent with the data shown in Fig. 6. Furthermore, quantum oscillation measurements[39–41] on the Nb and Cu homologues suggest that the planar cross-section of the Fermi surface shows little warping, indicating that effective mass anisotropy cannot be the sole cause of the anisotropy in $H_{c2}$.

However, for the more general case of anisotropic gaps the GL effective masses are given as[42]

$$\frac{1}{m_i} = \frac{1}{4\pi^3 \hbar N} \oint dS \phi^2(\boldsymbol{k}) \frac{v_i^2}{v_F} \tag{6}$$

Here, $dS$ denotes an integral over the Fermi surface, $N$ is the electron density, $v_i$ is the *i*-component of the Fermi velocity, and $v_F$ is the magnitude of the Fermi velocity, both in general **k**-dependent. ϕ(**k**) describes the anisotropy of the gap over the Fermi surface, normalized such that its Fermi surface average is unity. For instance, for a spherical Fermi surface (isotropic normal state electronic structure) and a model gap anisotropy of ϕ(k) = sin(θ) (corresponding to two point nodes on the *c*-axis) the $H_{c2}$-anisotropy for fields applied along the *c*-axis and for fields applied transverse is $1/\sqrt{2}$. We expect that nodal gap structures with different forms of ϕ(k) and gaps with deep minima will show similar qualitative behavior. Namely, that $H_{c2}$ measured along the line connecting the nodes (minima) is lower than in a transverse direction as discussed in more detail below.

The above analysis is based on the conventional GL-relations applicable for a system with a single-component order parameter. However, the nematic $E_u$ state that has been proposed as a possible explanation of the twofold anisotropy of superconducting properties is characterized by a two-component order parameter[11,30], which can be expressed as a linear superposition of the two basis gap functions $\Delta_{4x}$ and $\Delta_{4y}$. An in-depth analysis of the upper critical field of a superconductor with trigonal symmetry and two-component order parameter has been presented in Ref. 43. Three nematic domains related by rotations of 120° should arise in the sample, giving rise to overall threefold symmetry. Instead, the vast majority of reported data including those presented here reveal a simple twofold anisotropy indicative of a single nematic domain. The theoretical analysis[43] reveals that, in contrast to a single component order parameter, a two component order parameter couples linearly to strain fields as parameterized by a coefficient δ, and that such strain fields may serve to pin the nematic vector into a single domain. In particular, for sufficiently strong pinning δ and near $T_{c0}$, the two component order parameter is effectively reduced to a single component which for δ > 0 is approximately $\Delta_{4x}$ and for δ < 0 it is approximately $\Delta_{4y}$. Then a temperature independent anisotropy of the upper critical field of $H_{c2}^{(a)}/H_{c2}^{(a^*)} = \sqrt{(J_1 + J_4)/(J_1 - J_4)}$ for δ > 0 and $H_{c2}^{(a)}/H_{c2}^{(a^*)} = \sqrt{(J_1 - J_4)/(J_1 + J_4)}$ for δ < 0 is expected with an angular dependence that is given by the conventional form (Eq. 2). Here, $J_1$ and $J_4$ are coefficients of the gradient terms in the two-component GL free energy in the notation of Ref. 43. Thus, depending on the values of these coefficients, a sizable in-plane anisotropy of $H_{c2}$ can arise even when the electronic structure is essentially isotropic. In particular, our observation that $H_{c2}^{(a)} > H_{c2}^{(a^*)}$ implies that the nodal $\Delta_{4x}$ state is realized.

It has been reported that magnetic effects may play an important role in the formation of the superconducting state in $Bi_2Se_3$-derived superconductors, i.e., $Nb_xBi_2Se_3$[16,44]. We therefore explored the temperature dependence of the normal state magnetization of $Sr_{0.1}Bi_2Se_3$. Fig. 6(b) shows data for crystal #3 measured in a field of 1 T applied along various in-plane directions. Within the experimental uncertainties, the normal state magnetization is isotropic in the basal plane ruling out a magnetic origin of the observed in-plane anisotropy of the superconducting state. Furthermore, in its normal state, $Sr_{0.1}Bi_2Se_3$ is diamagnetic, approaching a volume susceptibility of −2·10⁻⁶ (CGS) at high temperature. The measured magnetic susceptibility, χ, contains several contributions[45], $\chi = \chi_{core} + \chi_P + \chi_L + \chi_{VV} + \chi_{CW}$. Here, $\chi_{core}$ represents the core diamagnetism, $\chi_P$ and $\chi_L$ the Pauli paramagnetism and Landau diamagnetism of the conduction electrons, respectively, $\chi_{VV}$ the van Vleck paramagnetism and $\chi_{CW}$ a Curie-Weiss contribution, possibly due to magnetic impurities. $\chi_{core}$ is temperature independent and isotropic, whereas $\chi_P$, $\chi_L$, and $\chi_{VV}$ are temperature independent but in general anisotropic, depending on the band structure and orbital structure. Since the charge count of $Sr_{0.1}Bi_2Se_3$ is much lower than that of typical metals we neglect $\chi_P$ and $\chi_L$. With the help of tabulated values[46], $\chi_{core}$ of $Sr_{0.1}Bi_2Se_3$ can be estimated as $-2.3 \cdot 10^{-6}$ (CGS). Thus, the observed isotropic diamagnetic response of $Sr_{0.1}Bi_2Se_3$ is in large part caused by its core diamagnetism, which mainly stems from the $Se^{2-}$ ions. The van Vleck contribution may account for the difference between the measured and expected diamagnetic signals, $\chi_{VV} \sim 0.3 \cdot 10^{-6}$ (CGS). In addition, superimposed onto the diamagnetic signal is a paramagnetic contribution, which approximately follows a Curie-Weiss dependence [Fig. 6(b)]. This contribution is also isotropic, and we attribute it to residual magnetic impurities.

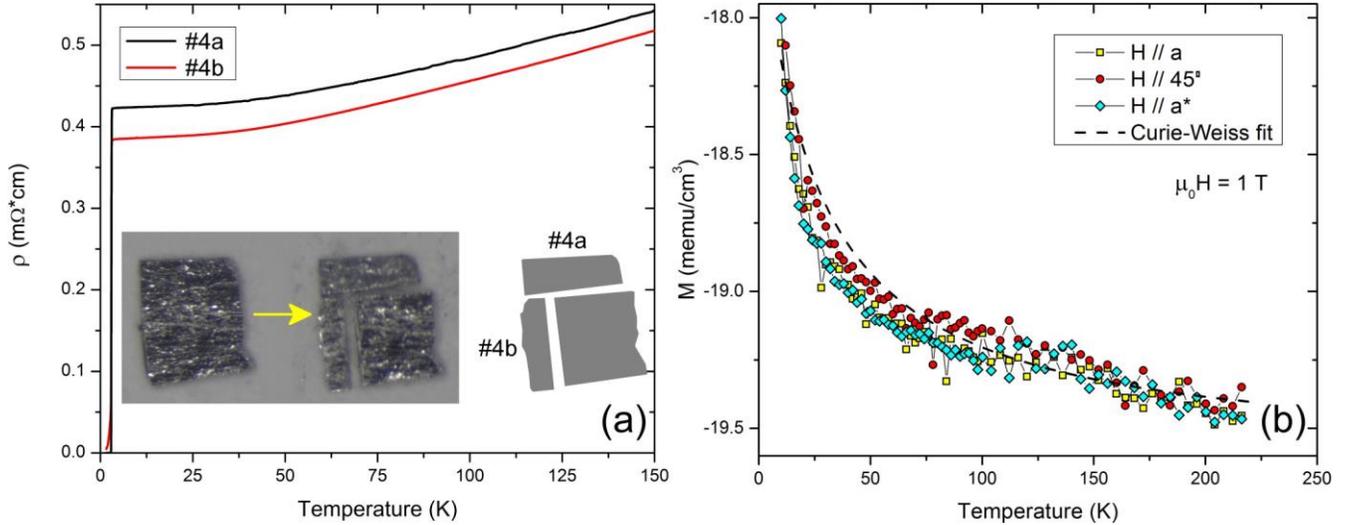

**Figure 6.** (a) Zero-field superconducting resistive transition ($T_c \approx$ 3.0 K) of $Sr_{0.1}Bi_2Se_3$ crystals #4a and #4b, cut at right angles to each other out of a single larger crystal. The inset shows a typical $\sim 1$ mm$^2$ crystal before and after cutting two transport samples out of it at right angles to each other. The anisotropy in $\rho$ is within the uncertainty of the dimensions of the electrical contacts. (b) Magnetization vs temperature of $Sr_{0.1}Bi_2Se_3$ crystal #3 with the field along the *a* axis (yellow), the *a\** axis (blue), and at 45° to either in the *aa\** plane (red). The material is diamagnetic, with a Curie-Weiss component (dashed line) possibly due to impurity contamination. The magnetization is essentially isotropic in-plane.

Deviations from the ideal $R\bar{3}m$ crystal symmetry have been proposed as possible causes of the twofold anisotropy itself or as mechanism of pinning the nematic vector into one domain. We performed x-ray diffraction studies on the crystals used here in order to search for these effects. These measurements revealed a high-degree of structural coherence and phase purity. The determined room-temperature lattice parameters are $a$ = 4.146 Å and $c$ = 28.664 Å, consistent with a rhombohedral $R\bar{3}m$ crystal symmetry derived from the Sr-intercalated $Bi_2Sr_3$ structure[15,36]. Figure 7 shows *l* scans centered on various [*h*,0,*l*] zones performed on the same crystal whose transport measurements are shown in Fig. 3 and Fig. 4. Multiple *h* values are shown; *h* = 0 (green circles), *h* = 1 (blue squares), *h* = 2 (pink triangles). At all (*h0l*) zones examined, only Bragg peaks for which 2*h*+*k*+*l* = 3*n* is satisfied are observed. This is the extinction rule for the $R\bar{3}m$ structure. The data shown in Fig. 7 reveal that this extinction rule is satisfied to a level of $10^{-6}$ implying that deviations from the ideal $R\bar{3}m$ structure are exceedingly small. Over the large illuminated area of the order of 0.4 x 0.4 mm$^2$, comparable to the sample size, there are three closely aligned grains with a mosaic of $\sim$ 0.04° each [see Fig. 7(b)], which is remarkable for a crystal formed from intercalating atoms between stacks of weakly coupled "quintuple layers". These measurements do not reveal, at room temperature, any crystal lattice distortions that could account for the large twofold anisotropy seen in the superconducting properties. It is unlikely, based on the smooth behavior observed in transport and magnetization data (Fig. 6) and calorimetry data[26], that there is any structural change at low temperature.

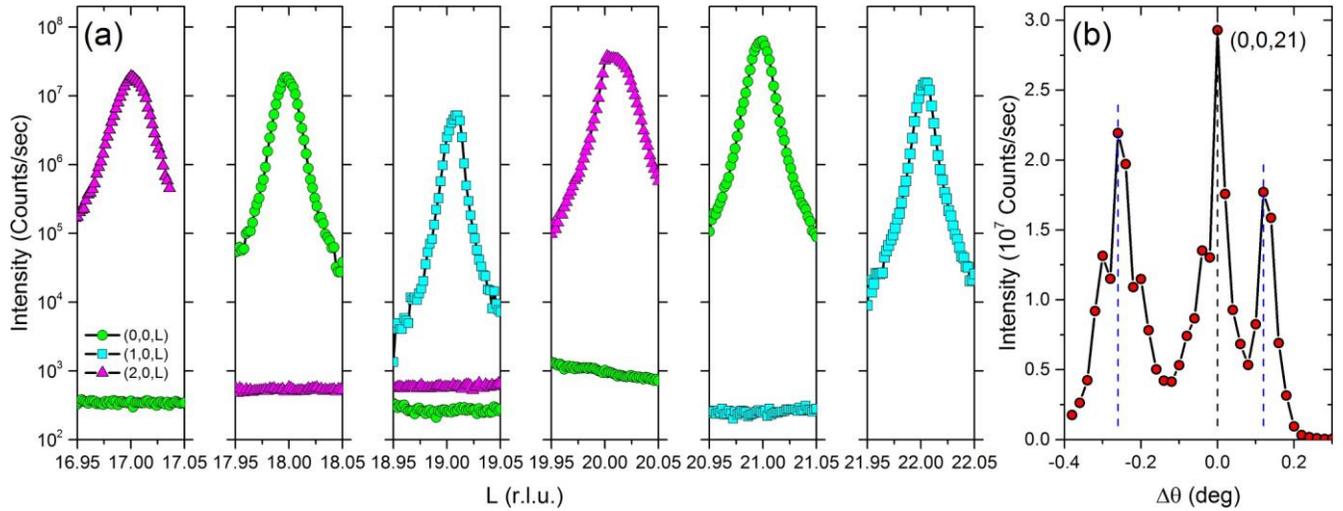

**Figure 7.** (a) Scans centered at ($h,0,l$) for multiple values of $l$ on $Sr_{0.1}Bi_2Se_3$ crystal #2 used for transport measurements. Multiple values of $h$ are shown; $h = 0$ (green circles), $h = 1$ (blue squares), $h = 2$ (pink triangles). The trigonal structure enforces an extinction rule unless $2h+k+l = 3n$, where $n$ is an integer. The allowed peaks show approximately 5 orders of magnitude more intensity than at $l$ values that are not allowed, showing the high quality of the crystal. Any distortions away from a perfect crystal structure would appear as violations of the extinction rule; none are seen. An additional crystal examined shows similar results. (b) Rocking curve centered at ($h,k,l$)=(0,0,21), showing three closely aligned major grains with a narrow mosaic spread of $\sim 0.04°$.

## Discussion

In addition to angular-dependent magnetotransport measurements we present the first thermodynamic determination of the anisotropy of the upper critical field of $Sr_{0.1}Bi_2Se_3$ crystals through measurements of the temperature dependence of the reversible magnetization. Both quantities yield a large twofold in-plane anisotropy of $H_{c2}$ in which the high-$H_{c2}$ direction is aligned with the $a$-axis of the crystal structure. Transport measurements on pairs of samples cut at 90° from the same starting crystal demonstrate that the in-plane anisotropy of $H_{c2}$ is tied to the crystal structure and is not induced by the current flow, consistent with the thermodynamic observations. These measurements also show that the normal state resistivity of $Sr_{0.1}Bi_2Se_3$ is isotropic in the plane, thereby excluding conventional effective mass anisotropy as a cause of the $H_{c2}$-anisotropy. Furthermore, temperature dependent measurements of the normal-state magnetization reveal that $Sr_{0.1}Bi_2Se_3$ is diamagnetic with an isotropic susceptibility of $\sim -2 \cdot 10^{-6}$ (CGS) which largely originates from the core diamagnetism. These results rule out a possible magnetic origin of the superconducting anisotropy. In addition, x-ray diffraction studies reveal a high degree of structural coherence and phase purity without any detectable deviations from the $R\bar{3}m$ crystal structure that could cause the twofold anisotropy. We thus conclude that the origin of the twofold anisotropy of the superconducting properties is likely caused by an anisotropic gap structure as realized in the nematic $E_u$ state. In fact, by specializing the general form of the GL free energy applicable to the two-component $E_u$ order parameter to the $\Delta_{4x}$ and $\Delta_{4y}$ basis functions, we retrieve an anisotropic single-component GL expression that can account for the experimental observations, and indicates the $\Delta_{4x}$ state is selected.

## Methods

Large high quality single crystals of $Sr_{0.1}Bi_2Se_3$ were grown by the melt-growth technique described in Ref. 37. All crystals regularly showed high volume fraction of superconductivity via magnetic susceptibility measurements with small variation in $T_c$ ranging from 2.9 K to 3.05 K. Thin crystals were cut from as-grown bulk crystals. The material cleaves easily in the basal plane yielding naturally flat surfaces parallel to $aa*$ in the lattice. Several crystal pairs were cut out of a larger piece in a mutually perpendicular arrangement as illustrated in the inset of Fig. 6(a). Some pairs were aligned parallel and perpendicular to the high-$H_{c2}$-direction [such as in Fig. 6(a)] whereas others were intentionally misaligned (such as in Fig. 2). Gold wires were then attached to the crystals using silver epoxy in a conventional 4-point measurement configuration. The crystals were mounted with their long axes parallel to each other such that the angle between current and applied in-plane magnetic field were always the same for both. An AMI 1 T superconducting vector magnet was used to apply magnetic field in arbitrary directions without having to physically rotate the sample, and currents smaller than or equal to 1 mA were used for the measurements. Slow rotation of the field direction in 2° increments ensured thermal equilibrium was maintained. The field was swept clockwise from 0° to 400° to eliminate any magnetic hysteresis effects. Magnetization measurements were performed in a 7T Quantum Design MPMS with samples mounted on a quartz glass fiber with GE varnish to minimize the background signal. X-ray measurements were performed at the 6-ID-B beamline at the Advanced Photon Source. A vertically focused x-ray beam of 8.979 keV was delivered to the sample. The sample was oriented such that measurements using a reflection geometry from a naturally cleaved surface normal to the *c*-axis can be carried out.

## Acknowledgements

Magnetization and magnetotransport measurements were supported by the U.S. Department of Energy, Office of Science, Basic Energy Sciences, Materials Sciences and Engineering Division. M.P.S. thanks ND Energy for supporting his research and professional development through the NDEnergy Postdoctoral Fellowship Program. K.W. acknowledges support through an Early Postdoc Mobility Fellowship of the Swiss National Science Foundation. G.D.G. was supported by the Office of Basic Energy Sciences, US Department of Energy under Contract de-sc0012704. R.D.Z. and J.A.S. were supported by the Center for Emergent Superconductivity, an Energy Frontier Research Center funded by US Department of Energy.

## Author contributions statement

G.G., J.A.S, R.D.Z. sample synthesis. K.W. sample fabrication, magnetotransport measurements, and x-ray diffraction. H.C. magnetization measurements. U.W. magnetization measurements and analysis. A.E.K. and K.W.S., theoretical analysis. Z.I. x-ray diffraction. W.-K.K. experiment design. M.P.S. magnetotransport measurements, analysis, manuscript writing with contributions of U.W. and K.W.. All authors reviewed the manuscript.

## Additional information

Competing financial interests: The authors declare no competing financial interests.

Supplemental Information

Superconducting and normal-state anisotropy of the doped topological insulator Sr$_{0.1}$Bi$_2$Se$_3$

M. P. Smylie,[1,2] K. Willa,[1] H. Claus,[1] A. E. Koshelev,[1] K. W. Song,[1] W.-K. Kwok,[1] Z. Islam,[3] G. D. Gu,[4] J. A. Schneeloch,[4,5] R. D. Zhong,[4,6] and U. Welp[1]

[1]Materials Science Division, Argonne National Laboratory, Argonne, IL 60439, USA
[2]Department of Physics, University of Notre Dame, Notre Dame, IN 46556, USA
[3]Advanced Photon Source, Argonne National Laboratory, Argonne, IL 60439, USA
[4]Condensed Matter and Materials Science Department, Brookhaven National Laboratory, Upton, NY 11793, USA
[5]Department of Physics and Astronomy, Stony Brook University, Stony Brook, NY 11794, USA
[6]Department of Materials Science and Engineering, Stony Brook University, Stony Brook, NY 11794, USA

In order to rule out doping variations as a cause of the difference in anisotropy seen using magnetotransport and magnetization measurements, we present determinations of the superconducting phase diagram of Sr$_{0.1}$Bi$_2$Se$_3$ crystal #5 using both techniques on the same sample.

Magnetoresistance:
Sr$_{0.1}$Bi$_2$Se$_3$ crystal #5 is bar-shaped, oriented such that the current flows along the a-direction. Using the 50% criterion, a zero-field T$_c$ of 3.04K is obtained for this sample. The resistive transitions in fields applied along the $a$, $a^*$ and $c$-directions are shown in Fig. S1. With increasing field, the transitions shift uniformly to lower temperature similar to those obtained on crystal #2 displayed in Fig. 3 of the article. The emergence of a substantial in-plane anisotropy is evident from a comparison of panels (a) and (b).

Magnetization:
The temperature dependence of the magnetization measured in several fields along the $a$, $a^*$ and $c$ directions and a direct comparison of the 4-kOe data for all three orientations are shown in Fig. S2. These data were taken on increasing temperature after cooling the sample in the indicated fields. Also included in Fig. S2(a) are the data in 2 kOe obtained on warming after the sample has been cooled in zero-field, showing that the magnetization is essentially reversible.
Near the transition, the magnetization is well described by a linear temperature dependence as indicated by the black lines. This dependence for the equilibrium magnetization is expected on the basis of Ginzburg-Landau theory and allows for a determination of $T_c(H)$ from the intersect with the $m$=0 line. The data in Fig. S2(d) clearly reveal the in-plane anisotropy in $T_c(H)$ and in the slopes of $m(T)$, in agreement with Fig. 5a in the article.

Phase diagram:
The phase boundaries deduced from the resistive 50% criterion are shown in Fig. S1(d) as solid circles. The general features of this phase diagram, namely the large in-plane anisotropy and a pronounced upward curvature near $T_c$, are similar to those in Fig. 3(d) of the article. The anisotropy deduced from these data decreases slightly with increasing temperature from Γ ~ 5.3 at 2.68 K to 5.15 at 2.8 K and 4.9 at 2.9 K. The shaded areas in Fig. S2(d) are bounded by the $H_{c2}$-lines obtained with the 2% and 98% criterion corresponding to values close to the zero-resistance point and the resistive onset, respectively. We note that the phase boundaries shift considerably depending on the criterion employed. However, the effect of the criterion is asymmetric for the $a$ and $a^*$ directions reflecting the observation that upon increasing field a tail starts developing for $H // a$, whereas for $H // a^*$ there appears rounding near the top of the transition. At the same time, the in-plane anisotropy changes from 3.1 for the 2% - criterion, to 5.15 at 50% and 5.0 at 98 %.
Fig. S2(d) also includes the $T_c(H)$ values obtained from the magnetization data. The error bars reflect the uncertainties in locating $T_c(H)$, see Figs. S2(a, b, c). In contrast to the resistive data, the magnetic data yield essentially linear phase boundaries with slopes of 24.6 kOe/K and 9.5 kOe/K for the *a* and *a\** directions, respectively, corresponding to an in-plane anisotropy of Γ ~ 2.6. The same value for the in-plane anisotropy is also obtained from the slopes of the *m*(*T*)-data in Fig. S2(d). We note that the magnetically determined phase boundaries for the *a\** and *c* directions almost coincide, in agreement with the resistive data. The unusual situation arises that the magnetically determined phase boundary lies above the resistive 50%-line for *H* // *a\** whereas it lies below for *H* // *a* implying that there is no criterion that would bring both data sets in alignment. The reasons for this unexpected behavior are not understood at present, and may be related to the unusual positive curvature of the resistively determined $H_{c2}$ observed in all samples, or to the existence of surface states, which may have different superconducting properties than the bulk.

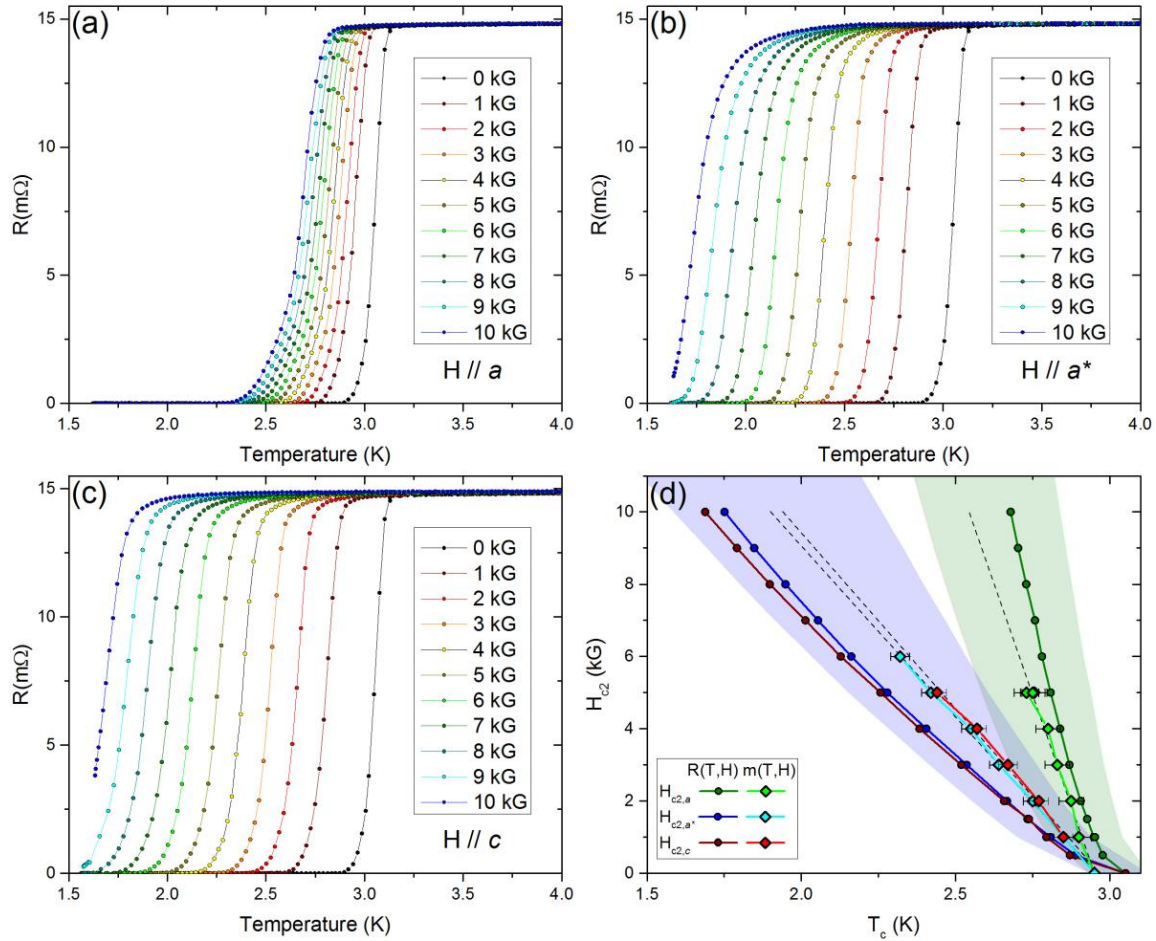

Fig. S1: Temperature dependence of the resistance of $Sr_{0.1}Bi_2Se_3$ crystal #5 measured in various magnetic fields applied along the principal crystal directions. (a) Field vector *H* // *a*. (b) Field vector *H* // *a\**. (c) Field vector *H* // *c*. (d) Superconducting phase diagram; circles are from R(T,H), diamonds are from m(T,H). The two shaded regions mark the phase boundaries for the resistive 2% and 98% criterion for the *a* (green) and *a\** (blue) directions. For each, the 2% criterion has the lower $H_{c2}$.

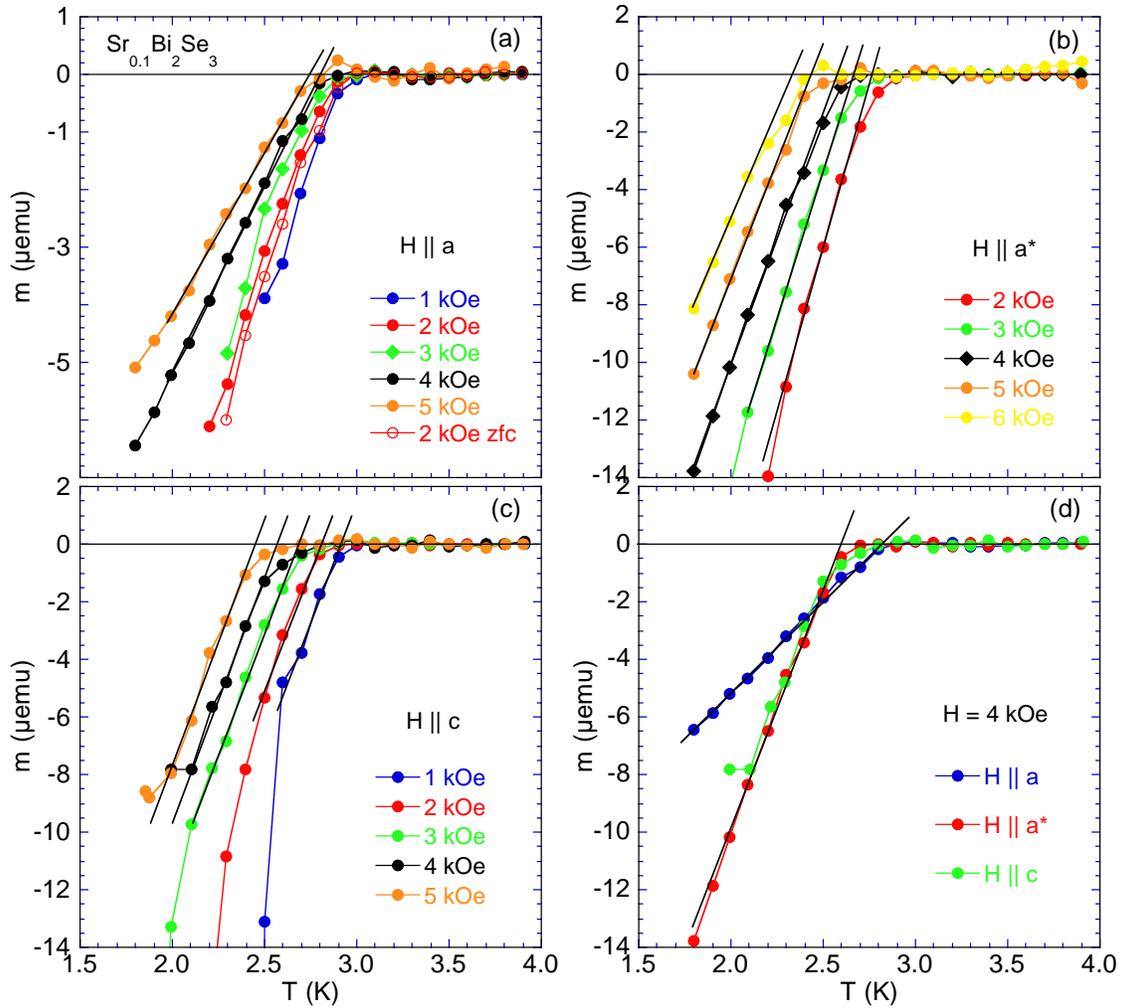

Fig. S2: Temperature dependence of the magnetic moment of $Sr_{0.1}Bi_2Se_3$ crystal #5 measured in various magnetic fields applied along the principal crystal directions. The black lines mark the almost linear temperature dependence of the magnetic moment near $T_c$. (a) Field vector $H \parallel a$. (b) Field vector $H \parallel a^*$. (c) Field vector $H \parallel c$. (d) Comparison of the 4-kOe data for all three orientations.